\documentclass[sigconf,screen]{acmart}

\AtBeginDocument{%
  }
\usepackage{cleveref}
\usepackage[autostyle, english = american]{csquotes}
\MakeOuterQuote{"}
\usepackage{hyperref}
\usepackage{booktabs}
\usepackage{graphicx}
\usepackage{threeparttable}

\copyrightyear{2025}
\acmYear{2025}
\setcopyright{acmlicensed}\acmConference[FSE Companion '25]{33rd ACM International Conference on the Foundations of Software Engineering}{June 23--28, 2025}{Trondheim, Norway}
\acmBooktitle{33rd ACM International Conference on the Foundations of Software Engineering (FSE Companion '25), June 23--28, 2025, Trondheim, Norway}
\acmDOI{10.1145/3696630.3730563}
\acmISBN{979-8-4007-1276-0/2025/06}

\begin{document}

\title{AI in Software Engineering: Perceived Roles and Their Impact on Adoption}

\author{Ilya Zakharov}
\email{ilia.zaharov@jetbrains.com}
\affiliation{%
  \institution{JetBrains Research}
  \city{Belgrade}
  \country{Serbia}
}

\author{Ekaterina Koshchenko}
\email{ekaterina.koshchenko@jetbrains.com}
\affiliation{%
  \institution{JetBrains Research}
  \city{Amsterdam}
  \country{Netherlands}
}

\author{Agnia Sergeyuk}
\email{agnia.sergeyuk@jetbrains.com}
\affiliation{%
    \institution{JetBrains Research}
    \city{Belgrade}
    \country{Serbia}
}

\begin{abstract}

This paper investigates how developers conceptualize AI-powered Development Tools and how these role attributions influence technology acceptance. Through qualitative analysis of 38 interviews and a quantitative survey with 102 participants, we identify two primary Mental Models: AI as an \textbf{inanimate tool} and AI as a \textbf{human-like teammate}. Factor analysis further groups AI roles into \textbf{Support Roles} (e.g., assistant, reference guide) and \textbf{Expert Roles} (e.g., advisor, problem solver). We find that assigning multiple roles to AI correlates positively with Perceived Usefulness and Perceived Ease of Use, indicating that diverse conceptualizations enhance AI adoption. These insights suggest that AI4SE tools should accommodate varying user expectations through adaptive design strategies that align with different Mental Models.

\end{abstract}



\maketitle

\section{Introduction}

Understanding how users conceptualize interactive systems is central to Human-Computer Interaction (HCI). The concept of \textbf{Mental Models} provides a lens through which researchers can study how individuals predict and interpret the behavior of complex technologies~\cite{staggers1993mental}. As Artificial Intelligence (AI) systems become increasingly integrated into everyday life, studying users' Mental Models of AI has become both more urgent and more challenging. Unlike traditional software, AI systems exhibit dynamic, probabilistic behaviors, making their internal logic less transparent to users. This ambiguity often leads users to engage in \textbf{Theory of Mind} reasoning --- attributing human-like intentions, emotions, or distinct roles to AI systems~\cite{dennett1989intentional}.

Such role attributions are not merely cognitive artifacts; they may have significant implications for how users interact with and adopt AI technologies. Prior research suggests that the way users frame AI's role in their workflows can influence trust, reliance, and overall adoption. The \textbf{Technology Acceptance Model (TAM)} has long identified \textbf{Perceived Usefulness (PU)} and \textbf{Perceived Ease of Use (PEU)} as key determinants of technology adoption~\cite{venkatesh2000theoretical}. PU reflects the belief that a system enhances performance, while PEU denotes how effortless users perceive it to be. Emerging evidence also suggests that how developers conceptualize their tools is closely related to their adoption~\cite{li2024ai,russo2024navigating,choudhuri2024guides}. However, it remains unclear whether and how the specific \textbf{roles} users attribute to AI systems shape these adoption factors. 

This paper investigates the relationship between \textbf{AI role attribution and technology acceptance in the realm of AI for Software Engineering (AI4SE)} through the following research questions:
\begin{itemize}
    \item \textbf{RQ1:} Which roles do users attribute to AI-powered Development Tools?
    \item \textbf{RQ2:} How do users qualitatively describe the roles they attribute to AI-powered Development Tools?
    \item \textbf{RQ3:} To what extent are the assigned roles of AI associated with technology acceptance, including Perceived Usefulness and Perceived Ease of Use?
\end{itemize}

By exploring these questions, this work aims to bridge the gap between theoretical models of user cognition and practical determinants of the adoption of AI-powered Development Tools. We argue that understanding how AI role attributions influence technology acceptance can inform both AI4SE design and broader discussions on human-AI collaboration.

\section{Background}

The adoption of AI4SE has given rise to a new class of AI-powered Development Tools designed to assist developers in various tasks, such as code completion, bug detection, and automated testing, \textit{e.g.} ~\cite{GitHubCopilot, CursorAI, TabnineAI}. How individuals will engage, predict, and interpret this technology would be dictated by their \textbf{Mental Models} ~\cite{staggers1993mental}, a mental representation of some system that supports understanding, reasoning, and prediction. They allow people to mentally simulate the behavior of a system. Understanding how developers conceptualize AI tools is essential for improving both tool design and user experience.

Alterations in a person's Mental Model of an AI may influence interactions with that system \cite{pataranutaporn_influencing_2023}.  Priming statements regarding the inner motives of conversational AI, categorized as caring, manipulative, or neutral affected ratings of trustworthiness, empathy, and performance. The most positive statements were associated with the caring motive. A feedback loop was also identified, wherein the user and AI mutually reinforced the user's Mental Model over a short period. These findings underscore the significance of how AI systems are introduced, as this can substantially affect user interactions and overall experiences.

Whether viewing AI tools as potentially having roles is the correct way to think about it is currently a matter for debate. There is research that shows that people might not actually attribute mind and mental states to AI tools \cite{astobiza2024do}. At the same time, another study showed thinking of AI as having some agency might be beneficial for its usage.  Using a factorial design, the study found that AI systems operating autonomously were rated as more trustworthy, and those using subjective language were perceived as both more trustworthy and likable. Importantly, the combination of autonomy and subjectivity resulted in the highest ratings for trust, likability, and overall quality. Additionally, participants' levels of AI literacy moderated the effects of source autonomy and language subjectivity on trust and chat quality evaluations. These findings offer valuable insights into factors that shape users' mental models and evaluations of AI behavior.

Given that humans naturally develop \textbf{Theories of Mind} to understand the intentions and behaviors of those they interact with~\cite{dennett1989intentional}, it is intuitive to extend this process to AI systems. In fact, attributing a "mind" to AI seems to be a natural extension of our innate cognitive processes. Moreover, this perspective can be framed as a mutual process --- where both humans and AI iteratively shape their mental models of one another. Some research even claims that the correct way to address human-computer interaction when talking about AI tools is according to the Mutual Theory of Mind framework \cite{wang2022mutual}. MToM highlights the dynamic, reciprocal nature of human-AI interactions and outlines three key components that iteratively influence communication between humans and AI: interpretation, feedback, and mutuality. Experimental studies in which researchers tried to simulate human-AI interaction according to this framework brought mixed results \cite{zhang2024mutual}. In a shared human-AI task, adding a MToM module did not significantly impact the team's objective performance but did reduce instances of cooperation failure.

Overall, existent research emphasizes that users' Mental Models and the way AI systems are framed can significantly shape trust, empathy, and overall evaluations of AI behavior. While some studies suggest that attributing a mind to AI may enhance user experience, contrasting evidence indicates that mental state attributions are not universally applied. 

\section{Method}

For the \textbf{qualitative} part of our research, we re-analyzed data from our recent empirical qualitative study of developers' perceptions of AI-powered Development Tools in Integrated Development Environments (IDEs) \cite{Sergeyuk2024HumanAI}. In the spring of 2024, we interviewed 38 programmers with varied experience in programming and with AI to understand their needs and challenges when using AI-powered Development Tools. During each interview, participants were asked to define AI, which was the focus of our present secondary analysis. During the interviews, participants discussed tasks AI had to perform well and technical requirements they valued, such as minimizing hallucinations, matching code style, and providing up-to-date information. We noticed different ways in which programmers conceptualize AI-powered Development Tools. The new insights, which we describe further in the paper, ultimately motivated this current research. The details on how interviews were conducted, as well as the data gathered, are available in the original paper~\cite{Sergeyuk2024HumanAI}

For the \textbf{quantitative} analysis of the roles people assign to AI-powered Development Tools, we conducted an online survey. The link to the survey was sent to a curated list of people who had given their consent to participate in user studies conducted by JetBrains, a software development company that makes tools for software developers. The study was carried out according to JetBrains' ethical standards, adhering to the values and guidelines outlined in the ICC/ESOMAR International Code~\cite{iccesomar}. In total, 102 participants fully completed the survey in January 2025.

The survey consisted of questions regarding the roles participants assign to AI-powered Development Tools, factors from the technology acceptance model (PU and PEU), and the number of AI tools people have tried for coding. The full questionnaire can be found in the supplementary materials\footnote{Survey available at \href{https://doi.org/10.5281/zenodo.14973853}{DOI: 10.5281/zenodo.14973853}.}. 

The roles suggested in the survey were based on the Developer Ecosystem Report 2024, the survey conducted yearly by JetBrains, which is designed to monitor the habits and attitudes of developers across various domains~\cite{JetBrains2024}. The available role options included: assistant, tool, reference guide, content generator, problem solver, collaborator, senior colleague, junior colleague, teacher, advisor, reviewer, copilot, and companion. Additionally, there was an option to specify your own role or select "None of the above". 

To capture participants' perceptions regarding technology use, we administered a Revised TAM Questionnaire that contains 12 items for two scales--- six measuring PU and six measuring PEU\cite{lewis2018comparison}.  Each scale score was obtained by averaging the relevant items, providing quantitative indices of participants' acceptance and usability judgments regarding AI-powered Development Tools.

To analyze the structure behind the roles people assign to AI, we used factor analysis. A factor model was fitted with the Python package "factor-analyzer" with a varimax rotation~\cite{factor_analyzer}. We used visual inspection of the scree plot to choose the optimal number of factors. 
After fitting the model, the factor loadings were extracted. Loadings exceeding a threshold of >0.4 were interpreted as significant~\cite{rogers2022best}.

To examine the pairwise relationships among the study variables, we computed Pearson correlation coefficients (\(r\)) and their associated \(p\)-values using Python's SciPy library~\cite{2020SciPy-NMeth}.

\section{Findings}

In our research, we were interested in understanding the relationship between \textbf{AI role attribution and technology acceptance in the realm of AI for Software Engineering (AI4SE)}. All anonymized data analyzed in this study is available upon request.

\subsection{Qualitative analysis of interviews}

According to our qualitative data from interviews~\cite{Sergeyuk2024HumanAI}, about 80\% (out of 38 professionals) referred to AI-powered Development Tools as a \textit{machine, software,} or \textit{tool}. Many saw it as a \textit{natural language processor} that understands user requests, directs them to the right algorithms, and then translates results back into natural language. Others viewed it as a sophisticated \textit{search engine}, or a blend of search and language processing. Another common view was that AI is a \textit{pattern recognizer} that automates tasks by applying learned patterns and rules.

Interestingly, 20\% of participants defined AI in human terms.
They used words like \textit{colleague}, \textit{companion}, or \textit{assistant}, as if AI were another team member. One participant said: \textit{``I'd want it to be like a colleague, someone who understands my work and offers guidance when I fall short.''} Less experienced programmers tended to describe AI as a \textit{teacher} or \textit{expert}, while more experienced professionals often compared it to an \textit{intern} or \textit{junior engineer} whose work still needs review.

These findings highlight an interesting difference in the conceptualization of AI: the animate versus inanimate perspective --- as a \textbf{tool} or \textbf{team member}. Both the \textit{tool} and \textit{team member} groups shared some opinions, but their expectations differed. 

Developers who view AI as an inanimate tool tend to enforce stricter technical standards, such as minimizing hallucinations, ensuring code style consistency, and providing up-to-date information. They are more likely to stop using AI if these expectations are not met.

In contrast, those who view AI as a human-like team member while also having certain technical standards appreciate its ability to speed up work and reduce tedious tasks. They tend to display greater tolerance for AI assistants' limitations, just as they do with human colleagues. 

These insights align with broader research on technology acceptance, where Mental Models and the roles assigned to AI are considered crucial in shaping adoption behaviors~\cite{bansal2019beyond}.

\subsection{Quantitative analysis of the online survey}
 
For our quantitative analysis, we were first interested in the distribution of the roles that developers assign to AI-powered Development Tools. Among the 102 participants in the study, only 3 did not have any coding experience, and 9 had less than a year of experience. Their jobs included management, digital design, online advertising, and molecular biology. All other respondents had at least one year of professional coding experience (12 people with 1-2 years, 21 people with 3-5 years, 19 people with 6-10 years, 14 people with 11-15 years, and 24+ with 16 and more years of experience). 

We found that "assistant" and "tool" are the most frequently assigned roles for AI (n = 70 and n = 69, correspondingly), suggesting that most developers view AI as either a helper or an instrument to facilitate coding tasks. Other popular roles include "reference guide", "content generator", and "problem solver", highlighting AI's perceived value in providing information, generating code, and resolving technical challenges. Fewer mentions of roles such as "teacher", "mentor", "senior colleague", or "junior colleague" suggest that some developers do personify AI as a peer, albeit less commonly. 

\begin{figure*}
    \centering
    \includegraphics[width=.65\textwidth]{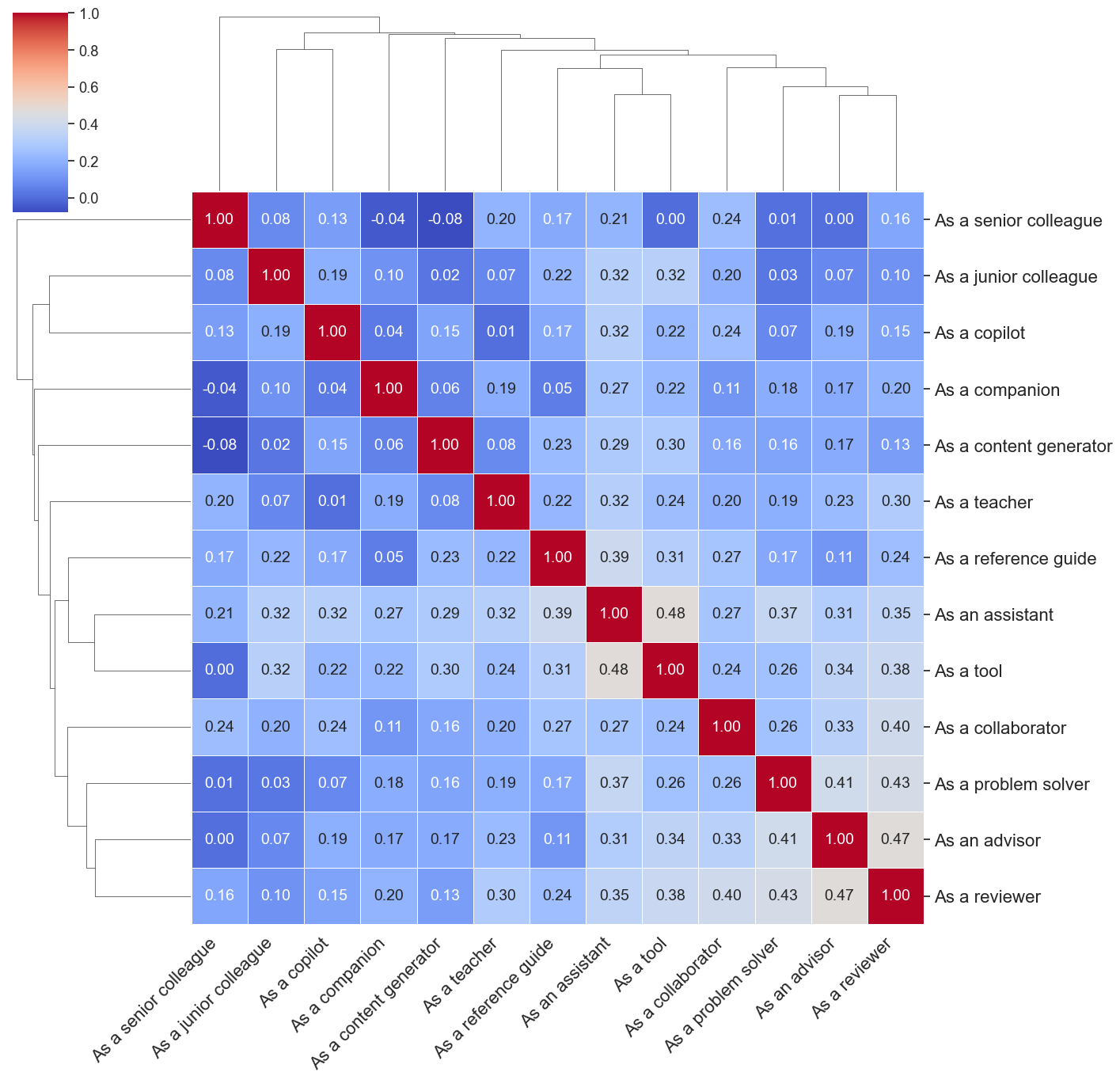}
    \caption{Hierarchical Clustering of the roles that developers assign to AI}
    \label{Figure 2}
\end{figure*}

The \Cref{Figure 2} illustrates the pairwise correlations among participants' perceptions of the roles of AI-powered Development Tools. The dendrogram with clustering on the rows and columns groups roles that frequently appear together. 

Several noteworthy patterns emerge: roles such as "assistant" and "tool" are close in the dendrogram, suggesting that participants who mention AI as an assistant often also view it as a tool. Conversely, more human-like roles such as "mentor", "teacher", or "colleague" cluster separately, indicating a distinct conceptualization of AI-powered Development Tools as a peer or guide. Overall, the heatmap underscores the existence of multiple Mental Models, ranging from a purely utilitarian view of AI to one in which AI is perceived more like a human collaborator.

For further analysis of the structure of AI roles, we performed factor analysis. The results of the scree plot showcased two factors underlying assigning different roles.

Significant factor loadings suggest that the roles people ascribe to AI can be broadly grouped into two dimensions. One factor tends to capture \textbf{Expert Roles} that emphasize evaluative, decision‐support, and expert functions \textit{e.g.,} high loadings for "As a problem solver" (loading = 0.599), "As an advisor" (loading = 0.662), and "As a reviewer" (loading = 0.691). Whereas the other factor gathers \textbf{Support Roles} that focus on supportive, assistive, and reference-type functions \textit{e.g.,} high loadings for "As an assistant" (loading = 0.657), "As a reference guide" (loading = 0.506), and "As a tool" (loading = 0.495).

Finally, we were interested in the relationship between AI roles and technology acceptance factors (PU, PEU and overall number of AI-powered Development Tools tried). The results are presented in \Cref{tab:correlation} 

\begin{table*}[h]
    \centering
    \small
    \begin{threeparttable}
   
    \resizebox{0.8\textwidth}{!}{%
        \begin{tabular}{lccccccc}
        \toprule
                           & AI roles & Expert Roles & Support Roles &    PU    &   PEU   & AI tools tried & Coding experience \\
        \midrule
        AI roles               & 1.0          & 0.7***         & 0.56***         & 0.59***  & 0.56*** & 0.28**    & -0.07               \\
        Expert roles           & 0.7***       & 1.0            & -0.15           & 0.41***  & 0.42*** & 0.19      & -0.20               \\
        Support roles          & 0.56***      & -0.15          & 1.0             & 0.34***  & 0.27**  & 0.19      & 0.18                \\
        PU                     & 0.59***      & 0.41***        & 0.34***         & 1.0      & 0.94*** & 0.33**   & -0.05               \\
        PEU                    & 0.56***      & 0.42***        & 0.27**          & 0.94***  & 1.0     & 0.34***  & -0.11               \\
        AT tools tried         & 0.28**       & 0.19           & 0.19            & 0.33**  & 0.34*** & 1.0      & 0.18                \\
        Coding experience      & -0.07        & -0.20          & 0.18            & -0.05   & -0.11   & 0.18     & 1.0                 \\
        \bottomrule
        \end{tabular}
    }
    \end{threeparttable}
    
    \begin{tablenotes}
    \centering
    \footnotesize
    \item \textit{Note:} Significance levels: *** $p < 0.001$, ** $p < 0.01$, * $p < 0.05$.
    \end{tablenotes}

    \caption{The relationship between AI roles, technology acceptance factors and coding experience}
    \label{tab:correlation}
\end{table*}

Both the total number of AI roles, as well as the Expert Roles and Support Roles factors, are positively correlated with technology acceptance factors (PU, PEU, and number of AI-powered Development Tools tried). Interestingly, the correlations between the developers' coding experience and other variables did not reach a statistically significant level in our data. 

\section{Discussion}

Our findings reveal that developers conceptualize AI-powered Development Tools along two primary dimensions: as \textbf{inanimate tools} or as \textbf{human-like team mates}. 
The hierarchical clustering and factor analysis further confirm that AI roles can be grouped into two broad categories: \textbf{Support Roles} (e.g., assistant, reference guide) and \textbf{Expert Roles} (e.g., advisor, problem solver). These findings suggest that developers who view AI as a tool enforce stricter performance expectations, whereas those who see AI as a collaborator tolerate imperfections, akin to how they interact with human colleagues.

The correlation analysis highlights a significant relationship between AI role attribution and Technology Acceptance. Both Support and Expert Role attributions positively correlate with \textbf{PU} and \textbf{PEU}, indicating that the more roles a developer assigns to AI, the more likely they are to find it useful and easy to integrate into their workflow. This suggests that Mental Models of AI directly influence technology adoption decisions.

Understanding how developers attribute roles to AI can inform the design of AI-powered Development Tools. Given that developers hold diverse expectations --- ranging from AI as a mere tool to AI as a teammate --- designers should consider personalization features that align with these varying Mental Models. For example, AI systems could offer configurable interaction styles, allowing users to choose between a "technical assistant" mode and a "collaborative partner" mode.

Additionally, since role attribution influences AI acceptance, onboarding strategies should explicitly shape how developers perceive AI tools. If an AI-powered coding assistant is positioned as a junior developer, users may apply appropriate oversight, whereas framing AI as an expert may lead to over-reliance. Such framing strategies should be carefully calibrated to align with AI's actual capabilities.

\section{Conclusion}

This study examined how developers conceptualize AI-powered Development Tools and how these role attributions influence technology acceptance. Our findings highlight the need for AI4SE tools to accommodate varying developer expectations.  Future research should explore how these conceptualizations evolve over time and how onboarding strategies can influence role attribution and trust in AI-powered Development Tools.

\bibliographystyle{ACM-Reference-Format}
\bibliography{References}


\begin{thebibliography}{21}


\ifx \showCODEN    \undefined \def \showCODEN     #1{\unskip}     \fi
\ifx \showISBNx    \undefined \def \showISBNx     #1{\unskip}     \fi
\ifx \showISBNxiii \undefined \def \showISBNxiii  #1{\unskip}     \fi
\ifx \showISSN     \undefined \def \showISSN      #1{\unskip}     \fi
\ifx \showLCCN     \undefined \def \showLCCN      #1{\unskip}     \fi
\ifx \shownote     \undefined \def \shownote      #1{#1}          \fi
\ifx \showarticletitle \undefined \def \showarticletitle #1{#1}   \fi
\ifx \showURL      \undefined \def \showURL       {\relax}        \fi
\providecommand\bibfield[2]{#2}
\providecommand\bibinfo[2]{#2}
\providecommand\natexlab[1]{#1}
\providecommand\showeprint[2][]{arXiv:#2}

\bibitem[Astobiza(2024)]%
        {astobiza2024do}
\bibfield{author}{\bibinfo{person}{Aníbal~M. Astobiza}.} \bibinfo{year}{2024}\natexlab{}.
\newblock \showarticletitle{Do people believe that machines have minds and free will? Empirical evidence on mind perception and autonomy in machines}.
\newblock \bibinfo{journal}{\emph{AI and Ethics}} \bibinfo{volume}{4}, \bibinfo{number}{4} (\bibinfo{year}{2024}), \bibinfo{pages}{1175--1183}.
\newblock


\bibitem[Bansal et~al\mbox{.}(2019)]%
        {bansal2019beyond}
\bibfield{author}{\bibinfo{person}{Gagan Bansal}, \bibinfo{person}{Besmira Nushi}, \bibinfo{person}{Ece Kamar}, \bibinfo{person}{Walter~S Lasecki}, \bibinfo{person}{Daniel~S Weld}, {and} \bibinfo{person}{Eric Horvitz}.} \bibinfo{year}{2019}\natexlab{}.
\newblock \showarticletitle{Beyond accuracy: The role of mental models in human-AI team performance}. In \bibinfo{booktitle}{\emph{Proceedings of the AAAI conference on human computation and crowdsourcing}}, Vol.~\bibinfo{volume}{7}. \bibinfo{pages}{2--11}.
\newblock


\bibitem[Biggs and Madnani(2024)]%
        {factor_analyzer}
\bibfield{author}{\bibinfo{person}{Jeremy Biggs} {and} \bibinfo{person}{Nitin Madnani}.} \bibinfo{year}{2024}\natexlab{}.
\newblock \bibinfo{title}{factor-analyzer: A Python package for factor analysis}.
\newblock \bibinfo{howpublished}{\url{https://github.com/EducationalTestingService/factor_analyzer}}.
\newblock
\newblock
\shownote{Version 0.5.1, accessed March 5, 2025}.


\bibitem[Choudhuri et~al\mbox{.}(2024)]%
        {choudhuri2024guides}
\bibfield{author}{\bibinfo{person}{Rudrajit Choudhuri}, \bibinfo{person}{Bianca Trinkenreich}, \bibinfo{person}{Rahul Pandita}, \bibinfo{person}{Eirini Kalliamvakou}, \bibinfo{person}{Igor Steinmacher}, \bibinfo{person}{Marco Gerosa}, \bibinfo{person}{Christopher Sanchez}, {and} \bibinfo{person}{Anita Sarma}.} \bibinfo{year}{2024}\natexlab{}.
\newblock \showarticletitle{What Guides Our Choices? Modeling Developers' Trust and Behavioral Intentions Towards GenAI}.
\newblock \bibinfo{journal}{\emph{arXiv preprint arXiv:2409.04099}} (\bibinfo{year}{2024}).
\newblock


\bibitem[Cursor(2025)]%
        {CursorAI}
\bibfield{author}{\bibinfo{person}{Cursor}.} \bibinfo{year}{2025}\natexlab{}.
\newblock \bibinfo{booktitle}{\emph{Cursor AI-powered code editor}}.
\newblock
\urldef\tempurl%
\url{https://www.cursor.com/}
\showURL{%
\tempurl}
\newblock
\shownote{Accessed: March 5, 2025}.


\bibitem[Dennett(1989)]%
        {dennett1989intentional}
\bibfield{author}{\bibinfo{person}{Daniel~C Dennett}.} \bibinfo{year}{1989}\natexlab{}.
\newblock \bibinfo{booktitle}{\emph{The intentional stance}}.
\newblock \bibinfo{publisher}{MIT press}.
\newblock


\bibitem[GitHub(2025)]%
        {GitHubCopilot}
\bibfield{author}{\bibinfo{person}{GitHub}.} \bibinfo{year}{2025}\natexlab{}.
\newblock \bibinfo{booktitle}{\emph{GitHub Copilot}}.
\newblock
\urldef\tempurl%
\url{https://github.com/features/copilot}
\showURL{%
\tempurl}
\newblock
\shownote{Accessed: March 5, 2025}.


\bibitem[ICC/ESOMAR({[n.\,d.]})]%
        {iccesomar}
\bibfield{author}{\bibinfo{person}{ICC/ESOMAR}.} \bibinfo{year}{[n.\,d.]}\natexlab{}.
\newblock \bibinfo{title}{{International Code on Market, Opinion and Social Research and Data Analytics}}.
\newblock \bibinfo{howpublished}{\url{https://esomar.org/uploads/attachments/ckqtawvjq00uukdtrhst5sk9u-iccesomar-international-code-english.pdf}}.
\newblock
\newblock
\shownote{{Accessed: October 2024}}.


\bibitem[JetBrains(2024)]%
        {JetBrains2024}
\bibfield{author}{\bibinfo{person}{JetBrains}.} \bibinfo{year}{2024}\natexlab{}.
\newblock \bibinfo{title}{The State of Developer Ecosystem 2024}.
\newblock
\urldef\tempurl%
\url{https://www.jetbrains.com/lp/devecosystem-2024/}
\showURL{%
\tempurl}
\newblock
\shownote{Accessed: February 13, 2025}.


\bibitem[Lewis(2018)]%
        {lewis2018comparison}
\bibfield{author}{\bibinfo{person}{James R.~(Jim) Lewis}.} \bibinfo{year}{2018}\natexlab{}.
\newblock \showarticletitle{Comparison of Four TAM Item Formats: Effect of Response Option Labels and Order}.
\newblock \bibinfo{journal}{\emph{UXPA Journal}} \bibinfo{volume}{14}, \bibinfo{number}{4} (\bibinfo{year}{2018}), \bibinfo{pages}{224--236}.
\newblock
\urldef\tempurl%
\url{https://uxpajournal.org/tam-formats-effect-response-labels-order/}
\showURL{%
\tempurl}


\bibitem[Li et~al\mbox{.}(2024)]%
        {li2024ai}
\bibfield{author}{\bibinfo{person}{Ze~Shi Li}, \bibinfo{person}{Nowshin~Nawar Arony}, \bibinfo{person}{Ahmed~Musa Awon}, \bibinfo{person}{Daniela Damian}, {and} \bibinfo{person}{Bowen Xu}.} \bibinfo{year}{2024}\natexlab{}.
\newblock \showarticletitle{AI tool use and adoption in software development by individuals and organizations: a grounded theory study}.
\newblock \bibinfo{journal}{\emph{arXiv preprint arXiv:2406.17325}} (\bibinfo{year}{2024}).
\newblock


\bibitem[Pataranutaporn et~al\mbox{.}(2023)]%
        {pataranutaporn_influencing_2023}
\bibfield{author}{\bibinfo{person}{Pat Pataranutaporn}, \bibinfo{person}{Ruby Liu}, \bibinfo{person}{Ed Finn}, {and} \bibinfo{person}{Pattie Maes}.} \bibinfo{year}{2023}\natexlab{}.
\newblock \showarticletitle{Influencing human–{AI} interaction by priming beliefs about {AI} can increase perceived trustworthiness, empathy and effectiveness}.
\newblock \bibinfo{journal}{\emph{Nature Machine Intelligence}} \bibinfo{volume}{5}, \bibinfo{number}{10} (\bibinfo{date}{Oct.} \bibinfo{year}{2023}), \bibinfo{pages}{1076--1086}.
\newblock
\showISSN{2522-5839}
\href{https://doi.org/10.1038/s42256-023-00720-7}{doi:\nolinkurl{10.1038/s42256-023-00720-7}}
\newblock
\shownote{Publisher: Nature Publishing Group}.


\bibitem[Rogers(2022)]%
        {rogers2022best}
\bibfield{author}{\bibinfo{person}{Pablo Rogers}.} \bibinfo{year}{2022}\natexlab{}.
\newblock \showarticletitle{Best practices for your exploratory factor analysis: A factor tutorial}.
\newblock \bibinfo{journal}{\emph{Revista de Administra{\c{c}}{\~a}o Contempor{\^a}nea}} \bibinfo{volume}{26}, \bibinfo{number}{06} (\bibinfo{year}{2022}), \bibinfo{pages}{e210085}.
\newblock


\bibitem[Russo(2024)]%
        {russo2024navigating}
\bibfield{author}{\bibinfo{person}{Daniel Russo}.} \bibinfo{year}{2024}\natexlab{}.
\newblock \showarticletitle{Navigating the complexity of generative ai adoption in software engineering}.
\newblock \bibinfo{journal}{\emph{ACM Transactions on Software Engineering and Methodology}} \bibinfo{volume}{33}, \bibinfo{number}{5} (\bibinfo{year}{2024}), \bibinfo{pages}{1--50}.
\newblock


\bibitem[Sergeyuk et~al\mbox{.}(2024)]%
        {Sergeyuk2024HumanAI}
\bibfield{author}{\bibinfo{person}{A. Sergeyuk}, \bibinfo{person}{E. Koshchenko}, \bibinfo{person}{I. Zakharov}, \bibinfo{person}{T. Bryksin}, {and} \bibinfo{person}{M. Izadi}.} \bibinfo{year}{2024}\natexlab{}.
\newblock \showarticletitle{The Design Space of in-IDE Human-AI Experience}.
\newblock \bibinfo{journal}{\emph{arXiv preprint}}  \bibinfo{volume}{arXiv:2410.08676} (\bibinfo{year}{2024}).
\newblock
\urldef\tempurl%
\url{https://arxiv.org/abs/2410.08676}
\showURL{%
\tempurl}


\bibitem[Staggers and Norcio(1993)]%
        {staggers1993mental}
\bibfield{author}{\bibinfo{person}{Nancy Staggers} {and} \bibinfo{person}{Anthony~F. Norcio}.} \bibinfo{year}{1993}\natexlab{}.
\newblock \showarticletitle{Mental models: concepts for human-computer interaction research}.
\newblock \bibinfo{journal}{\emph{International Journal of Man-machine studies}} \bibinfo{volume}{38}, \bibinfo{number}{4} (\bibinfo{year}{1993}), \bibinfo{pages}{587--605}.
\newblock


\bibitem[Tabnine(2025)]%
        {TabnineAI}
\bibfield{author}{\bibinfo{person}{Tabnine}.} \bibinfo{year}{2025}\natexlab{}.
\newblock \bibinfo{booktitle}{\emph{Tabnine AI Code Completion}}.
\newblock
\urldef\tempurl%
\url{https://www.tabnine.com/}
\showURL{%
\tempurl}
\newblock
\shownote{Accessed: March 5, 2025}.


\bibitem[Venkatesh and Davis(2000)]%
        {venkatesh2000theoretical}
\bibfield{author}{\bibinfo{person}{Viswanath Venkatesh} {and} \bibinfo{person}{Fred~D Davis}.} \bibinfo{year}{2000}\natexlab{}.
\newblock \showarticletitle{A theoretical extension of the technology acceptance model: Four longitudinal field studies}.
\newblock \bibinfo{journal}{\emph{Management science}} \bibinfo{volume}{46}, \bibinfo{number}{2} (\bibinfo{year}{2000}), \bibinfo{pages}{186--204}.
\newblock


\bibitem[Virtanen et~al\mbox{.}(2020)]%
        {2020SciPy-NMeth}
\bibfield{author}{\bibinfo{person}{Pauli Virtanen}, \bibinfo{person}{Ralf Gommers}, \bibinfo{person}{Travis~E. Oliphant}, \bibinfo{person}{Matt Haberland}, \bibinfo{person}{Tyler Reddy}, \bibinfo{person}{David Cournapeau}, \bibinfo{person}{Evgeni Burovski}, \bibinfo{person}{Pearu Peterson}, \bibinfo{person}{Warren Weckesser}, \bibinfo{person}{Jonathan Bright}, \bibinfo{person}{St{\'e}fan~J. van~der Walt}, \bibinfo{person}{Matthew Brett}, \bibinfo{person}{Joshua Wilson}, \bibinfo{person}{K.~Jarrod Millman}, \bibinfo{person}{Nikolay Mayorov}, \bibinfo{person}{Andrew R.~J. Nelson}, \bibinfo{person}{Eric Jones}, \bibinfo{person}{Robert Kern}, \bibinfo{person}{Eric Larson}, \bibinfo{person}{C~J Carey}, \bibinfo{person}{{\.I}lhan Polat}, \bibinfo{person}{Yu Feng}, \bibinfo{person}{Eric~W. Moore}, \bibinfo{person}{Jake VanderPlas}, \bibinfo{person}{Denis Laxalde}, \bibinfo{person}{Josef Perktold}, \bibinfo{person}{Robert Cimrman}, \bibinfo{person}{Ian Henriksen}, \bibinfo{person}{E.~A. Quintero},
  \bibinfo{person}{Charles~R. Harris}, \bibinfo{person}{Anne~M. Archibald}, \bibinfo{person}{Ant{\^o}nio~H. Ribeiro}, \bibinfo{person}{Fabian Pedregosa}, \bibinfo{person}{Paul van Mulbregt}, {and} \bibinfo{person}{SciPy~1.0 Contributors}.} \bibinfo{year}{2020}\natexlab{}.
\newblock \showarticletitle{{SciPy} 1.0: Fundamental Algorithms for Scientific Computing in Python}.
\newblock \bibinfo{journal}{\emph{Nature Methods}}  \bibinfo{volume}{17} (\bibinfo{year}{2020}), \bibinfo{pages}{261--272}.
\newblock
\href{https://doi.org/10.1038/s41592-019-0686-2}{doi:\nolinkurl{10.1038/s41592-019-0686-2}}


\bibitem[Wang and Goel(2022)]%
        {wang2022mutual}
\bibfield{author}{\bibinfo{person}{Qiaosi Wang} {and} \bibinfo{person}{Ashok~K. Goel}.} \bibinfo{year}{2022}\natexlab{}.
\newblock \showarticletitle{Mutual theory of mind for human-AI communication}.
\newblock \bibinfo{journal}{\emph{arXiv preprint arXiv:2210.03842}} (\bibinfo{year}{2022}).
\newblock


\bibitem[Zhang et~al\mbox{.}(2024)]%
        {zhang2024mutual}
\bibfield{author}{\bibinfo{person}{Shao Zhang}, \bibinfo{person}{Xihuai Wang}, \bibinfo{person}{Wenhao Zhang}, \bibinfo{person}{Yongshan Chen}, \bibinfo{person}{Landi Gao}, \bibinfo{person}{Dakuo Wang}, \bibinfo{person}{Weinan Zhang}, \bibinfo{person}{Xinbing Wang}, {and} \bibinfo{person}{Ying Wen}.} \bibinfo{year}{2024}\natexlab{}.
\newblock \showarticletitle{Mutual theory of mind in human-ai collaboration: An empirical study with llm-driven ai agents in a real-time shared workspace task}.
\newblock \bibinfo{journal}{\emph{arXiv preprint arXiv:2409.08811}} (\bibinfo{year}{2024}).
\newblock


\end{thebibliography}

\end{document}